# Quantum Tunnelling and Room-Temperature Superconductivity of Hydride from Size Effects

Xiaozhi Hu

Department of Mechanical Engineering

University of Western Australia

Perth, WA 6009, Australia

xiao.zhi.hu@uwa.edu.au

Superconductivity of a micron-sized hydride sample measured between metal probes under extreme pressure could be considered as a macroscopic quantum tunnelling phenomenon through metal-hydride-metal. The energy barrier height of hydride is regulated by pressure. The energy barrier width between tips of the metal probes should be minimized to limit the chance of exponential decay in electron tunnelling. There is also a thickness effect since thinner hydride samples around 1 micron are favoured for achieving higher superconductive temperatures. Hence, reduction in both barrier width and sample thickness is recommended to ensure optimum quantum tunnelling for realization of the room temperature superconductivity.

**Keywords:** Quantum Tunnelling; Room-Temperature Superconductivity (RT-S); Size Effects

A recent perspective on room-temperature superconductivity (RT-S) from a robust team of 16 scientists has reviewed over 180 publications [1], showing in general the superconductivity research has focused exclusively on superconductors themselves, their compositions, crystal structures, processing conditions, and key physical and electrical properties. One of the fresh and critical views emerged from the perspective is the reflection on potential benefits of "incorporating insights from fields beyond physics and materials science", which could be pivotal for future superconductivity research.

To explore potential benefits of different methodologies, this study examines room-temperature superconductivity (RT-S) of hydrides from the viewpoint of macroscopic quantum tunnelling through metal-hydride-metal, as in Figure 1(a). There are two clear advantages for considering the superconductivity of a micron-sized hydride sample between metal probes as a macroscopic quantum tunnelling phenomenon. First, it is conceivable that the energy barrier height of hydride is regulated by the pressure $P$ in a diamond anvil cell (DAC) together with the diamond tip topography at the centre and hydride sample thickness [2,3]. The energy barrier height will be decreased with increasing pressure in DAC so that the hydride sample will most likely go through the transformation from an insulator to a semi-conductor, then to a conductor and finally to a superconductor in a given temperature environment. Extreme pressure is essential for lowering the energy barrier height or creating a favourable quantum state in hydride for optimum quantum tunnelling or the free flow of Cooper pairs [4].

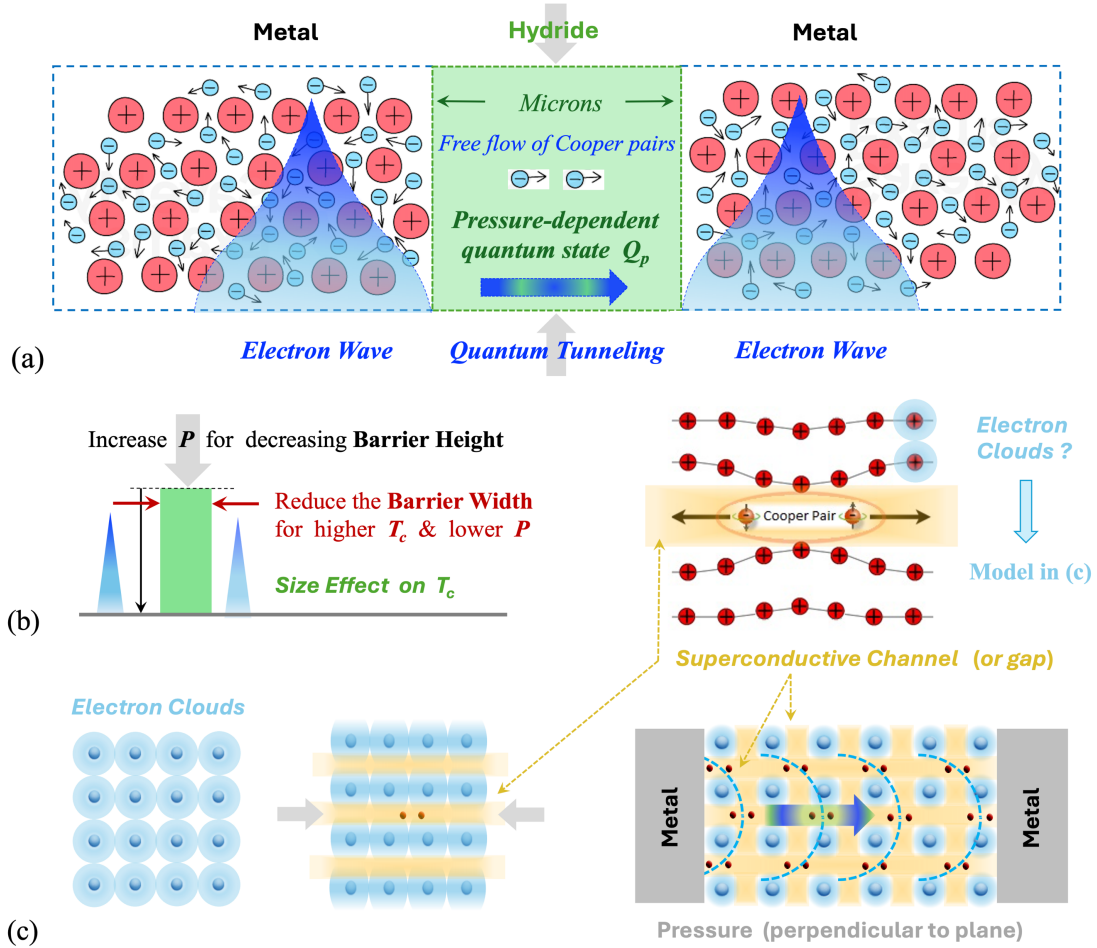

FIG. 1. (a) Quantum tunnelling through micron-sized hydride in the metal-hydride-metal system. Electron tunnelling is controlled by the quantum states of hydride and hydrogen atoms under extreme pressure. (b) Both superconductive temperature $T_c$ and pressure $P$ are strongly influenced by the barrier width. An "electron-free" positive lattice model is commonly used to illustrate the free flow of Cooper pairs. (c) "Low-electron-density" superconductive channels are generated under pressure through atom deformation and electron redistributions in the modified lattice containing electron clouds. Multiple superconductive channels are perpendicular in (a), connecting two metal sides. Formation and widening of superconductive channels with increasing pressure are responsible for the decrease in barrier height (more details in Figure 2).

Second, it is logical to minimize the energy barrier width, as depicted in Figure 1(b) to limit the chance of exponential decay in electron tunnelling. In this case, the barrier width is the planar section of the thin hydride sample between metal probes, typically around 5 microns across. The barrier width or distance between the tips of those metal probes should be minimized for higher $T_c$ measurements, considering the experimentally confirmed $T_c$ of $LaH_{10}$ is already around 250-260 K [5,6]. This "size effect" on the barrier width (not considered before) is critical to RT-S as it can be used either to reduce the pressure $P$ for a given $T_c$ or to increase $T_c$ for a fixed pressure $P$.

Classic quantum tunnelling deals with partial electron tunnelling through a nano-sized thin insulator [7,8]. Additionally, macroscopic quantum tunnelling has been studied through a current-biased Josephson tunnel junction in an electric circuit [9]. Metal-hydride-metal, as part of an electric circuit, is a new variant of macroscopic quantum tunnelling. The difference is superconductivity of a micron-sized hydride demands "electron-collision-free" quantum tunnelling. Clearly, both barrier height and width need to be minimised for optimum quantum tunnelling – the free flow of Cooper pairs [4].

It is widely recognized that the free flow of Cooper pairs through crystal lattices without electron collision is the fundamental mechanism of superconductivity [4]. Regardless of how unrealistic it may be, an "electron-free" positive lattice model as depicted in Figure 1(b) has been commonly used to illustrate the free flow of Cooper pairs. A lattice model with positive nuclei surrounded by electron clouds, as in Figure 1(c), would be more realistic. It is somewhat unexpected that the superconductive channels similar to those in the idealised "electron-free" positive lattice model in Figure 1(b) can be generated from the more realistic lattice model full of electrons as long as the applied pressure is sufficiently high to induce favourable electron redistributions from atom deformation (more details in Figure 2).

Instead of complete "electron-free", "low-electron-density" superconductive channels can be created under extreme uniaxial pressure. Formation and widening of the "low-electron-density" superconductive channels in the modified lattice model in Figure 1(c), akin to those "electron-free" channels in Figure 1(b), will depend on the applied pressure $P$ and the atomic number $Z$ of the element under consideration. Here, $Z$ = number of electrons that involve in the pressure-induced electron redistribution. For hydrogen-rich hydrides, the effective atomic number $Z = 1$ should be a good approximation. Clearly, hydrogen atoms with $Z = 1$ are most sensitive to the pressure-induced electron redistribution from atom deformation. Also, the local electron density in the "low-electron-density" superconductive channels in Figure 1(c) will be further reduced with increasing pressure $P$. In other words, the modified lattice model in Figure 1(c) with electron clouds can approach to the idea "electron-free" lattice model in Figure 1(b) under extreme pressure.

The unique quantum state of "deformed" atoms under extreme uniaxial pressure, as depicted in Figure 2, will lead to the desired electron redistributions and formation of the "low-electron-density" superconductive channels. The change in inter-atomic spacing under tension or compression as illustrated in Figure 2(a) is well accepted, which is commonly and phenomenologically elucidated by either the ball-and-spring model or the classic Lennard Jones Potential [10]. Physically, the spring effects against tension and compression can only be generated by electron redistributions from "atom deformation" [11]. Mathematically, the critical tensile strain $\varepsilon_c$ at the tensile fracture is fully defined by those physical properties in Figure 2(a) [12].

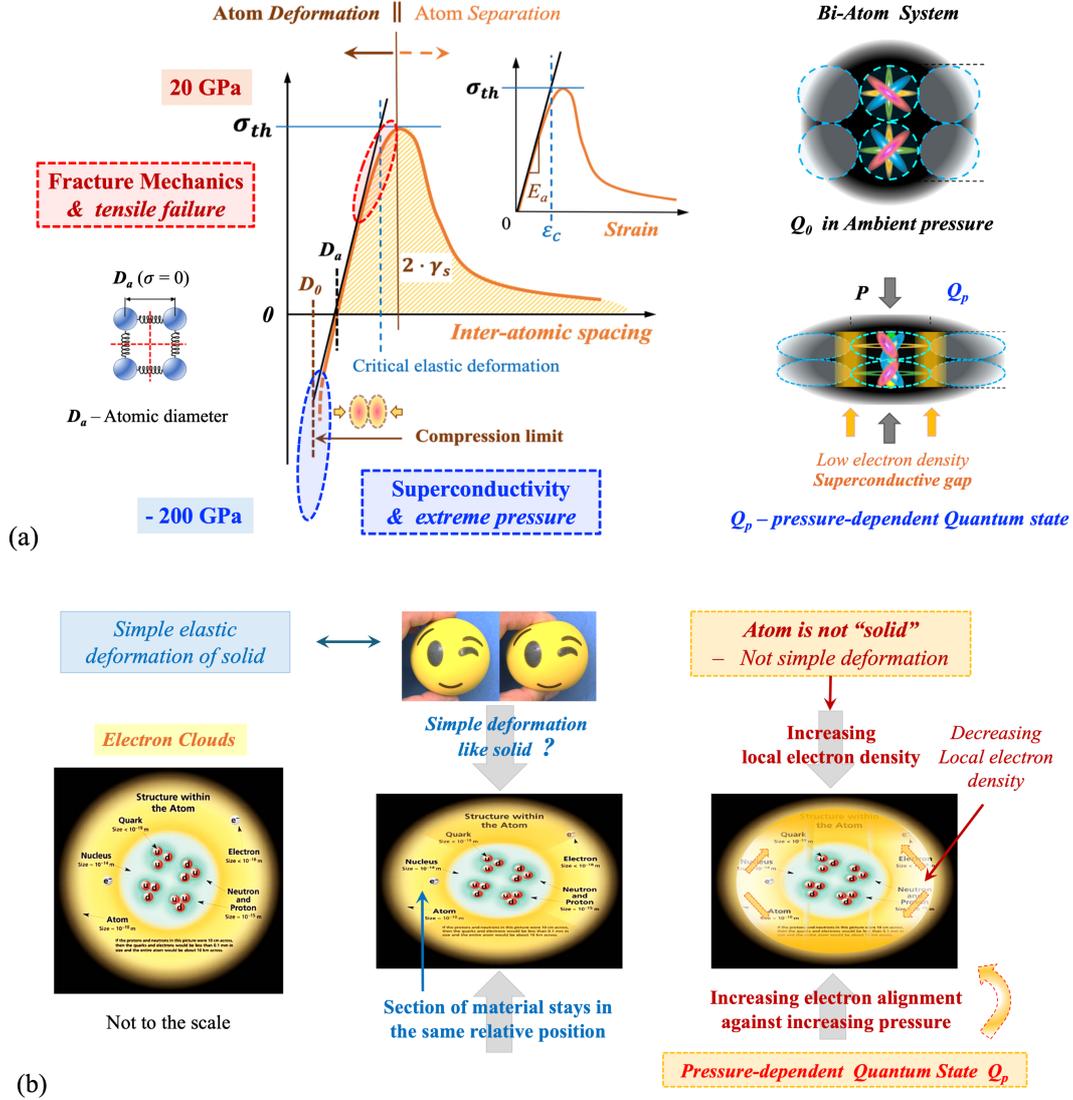

FIG. 2. (a) Well-studied tensile fracture at around 20 GPa and uncharted atom deformation under extreme pressure around - 200 GPa. (b) Atom is not solid, change in the inter-atomic spacing or atom deformation under extreme pressure will generate a unique pressure-dependent quantum state $Q_p$ [2,3,12]. Local electron distributions will be more aligned along the loading line by pulling electron waves away from the edge towards the centre, generating stronger repulsion between atoms. The pressure-dependent quantum state $Q_p$ of hydrogen atoms determines the quantum states of hydrogen-rich hydrides and their superconductivity.

Contrary to the well-defined tensile fracture at around 20 GPa, superconductivity-related atom deformation under extreme compression around - 200 GPa is in an uncharted territory, with only the non-linear modulus $E_{a\text{-}comp}$ and the possible compression limit $D_0$. The critical tensile strain $\varepsilon_c$ is linked to the theoretical strength $\sigma_{th}$, elastic modulus $E_a$ at the atomic scale, the surface energy $\gamma$, atomic diameter $D_a$ and Poisson's ratio $v$ [12].

$$\varepsilon_c = \frac{\sigma_{th}}{E_a} = \frac{1}{6 \cdot (1-v^2)} \cdot \frac{\gamma_s}{\sigma_{th}} \cdot \frac{1}{D_a} \qquad (1)$$

These well-defined physical properties may provide useful indications for the compressive atom deformation under extreme compression. Under extreme pressure $P$ around - 200 GPa, the modulus $E_{a\text{-}comp}$ will be higher than $E_a$ in tension. Since the compressive strain variation is too small, the compressive stress should be used and tentatively linked to the physical properties under tension, i.e.,

$$E_{a-comp} \cdot \varepsilon_{comp} \propto \sigma_{comp} \propto \frac{E_{a-comp}}{6 \cdot (1-v^2)} \cdot \frac{\gamma_s}{\sigma_{th}} \cdot \frac{1}{D_a} \propto P \qquad (2)$$

Electron redistributions and alignment are indirectly reflected through the modulus $E_a$ and $E_{a\text{-}comp}$.

Since atoms are not solid but formed by fast moving electron waves, atom deformation under extreme pressure illustrated in Figure 2(b) is not simple solid deformation but involving electron redistributions for generation of the natural repulsion against further compression. Atom deformation around - 200 GPa, pertinent to pressure-induced superconductivity of hydrogen-rich hydrides, is unique and governed by the pressure-dependent quantum state $Q_p$ of hydrogen atoms. That is electron waves and distributions in hydrogen atoms under - 200 GPa pressure are completely different to those of hydrogen atoms in the ambient pressure. Atom deformation and electron redistribution under pressure emulate variations in the pressure-dependent charge density wave and Fermi surface [13].

To generate the increasing repulsion against extreme pressure, as illustrated in Figure 2(a), it is assumed in this study that the local electron density along the loading line has to be increased. Consequently, a "low-electron-density" gap is created between atoms in the lateral direction, and then "low-electron-density" superconductive channels in the scale of lattice structures are created. That is the uniqueness of hydrogen atom deformation under - 200 GPa pressure creates the "low-electron-density" superconductive channels illustrated in Figure 1. Clearly, superconductivity of hydrides cannot be properly explained if the quantum state of hydrogen atoms under extreme pressure is not fully understood.

The concept of superconductive channel may even be useful for 2D lattice materials. A narrow "electron-free or low electron-density" superconductive channel is formed by placing two 2D lattice sheets together as illustrated in Figure 3(a). A "low electron-density" superconductive channel may be more realistic for 2D lattices, considering the Schrodinger quantum atomic model [14]. Changing the stacking orientation or position may create a broader channel (larger effective channel area) depending on the surface topography of a 2D lattice, e.g., the magic angle of two graphene sheets [15] and specific angles of twisted bilayer $WSe_2$ [16]. Neglecting the atomic scale variation or zigzag superconductive path, a 2D superconductive channel could be the common feature in various superconductors. A few 2D-like crystal structures of various superconductors have been summarised in [1], including interfaces, low-dimensionality, doping and intercalation. Within those 2D interfacial regions, electron distributions have been modified either chemically or by residual stress or strain linked to the material compositions, processing and post-processing treatment conditions, e.g., "pressure quenching" [1] and "strain-engineering" [17]. The interplay between atomic and electronic structures of $La_3Ni_2O_7$ thin film superconductors is emphasized in [17]. Likewise, Figure 1 and Figure 2 summarize the interplay between atom deformation, electron redistribution and superconductive channels in hydrogen-rich hydrides.

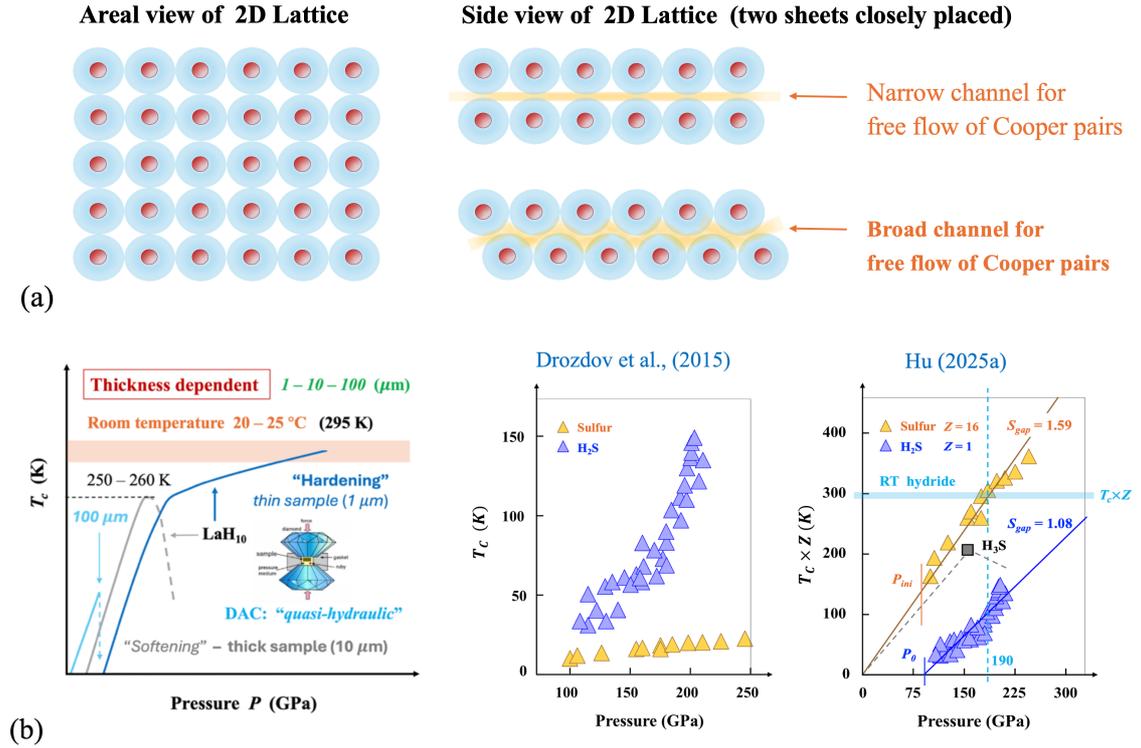

FIG. 3. (a) Areal view of a 2D lattice. A narrow "electron-free" superconductive channel is formed by placing two 2D lattice sheets closely together. Changing the stacking orientation or position may create a broader channel (wider and more effective channel cross-section) depending on the surface topography of the 2D lattice. (b) Initial linear $T_c – P$ relation is followed by either "softening" or "hardening", depending on the sample thickness. Comparison of the original $T_c – P$ measurements [18] and the $T_c – P/Z$ relations [2]. Ultra-thin hydride samples (around 1 micron) with the "hardening" behaviour are preferred for higher $T_c$ measurements [3].

In the case of hydrogen-rich hydrides, applying pressure is more effective to create the superconductive channels through electron redistributions and atom deformation. Considering the hydrides, the 2D case in Figure 3(a), and 2D-like crystal structures of various superconductors summarised in the recent perspective [1] and in [15,16], it seems the formation of "low-electron-density" superconductive channels could be the core mechanism for many superconductors although the means to achieve such favourable quantum states are different.

The concept of superconductive channels in Figure 1(b,c) can be adopted to derive an approximate relation between the superconductive temperature $T_c$ and pressure $P$ for various superconductors. Many experimental results show an approximate linear $T_c$ - $P$ relation exists in the initial increasing phase [2]. For instance, the experimental measurements of sulfur and sulfur hydride [18], shown in Figure 3(b), can be approximately described by linear $T_c$ - $P$ relations. Here, we are going to show the superconductive behaviours after the initial linear $T_c$ - $P$ relation reveals the important thickness effect on $T_c$ measurements, which can be critical to RT-S. The sample thickness effect could generate additional 15% enhancement in $T_c$, which is sufficient to raise the current experimentally-confirmed 250-260 K to 288-299 K or room temperature range [2,3].

Considering the unique characteristics of atom deformation and the pressure-dependent quantum state $Q_p$ in Figure 2(b), it can be assumed approximately that the width of "low-electron-density" superconductive channel $S_c$ between atoms in the idealized lattice model in Figure 1(b,c) is proportional to the pressure $P$, i.e.,

$$S_c \propto P \tag{3}$$

The compressive strain indicated approximately by Eq. (2) could provide the support for the lateral deformation in Eq. (3). The "low-electron-density" superconductive channel width $S_c$ in Figure 1(b,c) will be more difficult to generate if atoms contain a larger number of electrons or the element has a higher atomic number $Z$. Approximately, it can be assumed that:

$$S_c \propto 1/Z \tag{4}$$

For a given pressure $P$ and a given material element or compound with a specific atomic number $Z$, the superconductive temperature $T_c$ is approximately proportional to the "low-electron-density" superconductive channel width $S_c$, i.e.,

$$T_c \propto S_c \tag{5}$$

From the approximate relations in Eq. (3) to (5), it can be written that:

$$T_c \propto P/Z \tag{6a}$$

or simply,

$$T_c \propto P \tag{6b}$$

Eq. (6b) is for a given superconductor or compound with an equivalent atomic number $Z$. For hydrogen-rich hydrides, $Z = 1$ should be a good approximation.

Different superconductors can have different starting points in their $T_c$ - $P$ relations when the pressure $P = 0$. It is possible that for $P = 0$, the superconductive temperature $T_c = T_0 \geq 0$. It is also possible that for $T_c = 0$, $P = P_0 \geq 0$. So, in general, Eq. (6) can be rewritten as follows:

$$(T_c - T_0) \cdot Z = S_{gap} \cdot (P - P_0) \tag{7}$$

Here $S_{gap}$ is the slope of the $T_c$ - $P$ relation linked to the superconductive channels in Figure 1(c).

The linear relation described in Eq. (7) is only approximate since it is not obtained through rigorous mathematical derivations. Yet, it does describe well the initial $T_c$ - $P$ relations for many different superconductors [2], including those in Figure 3(b) and Figure 4. The $T_c \times Z$ - $P$ relations in Figure 3(b) based on Eq. (7) have somewhat connected the sulfur and sulfur hydride results together. $H_3S$ (with "softening" characteristics) is closer to the sulfur relation than $H_2S$ due to its higher hydrogen content.

One unexpected after-effect of this phenomenological $T_c \times Z$ - $P$ relation turns out to be critical to RT-S measurements. After the initial linear phase of $T_c \times Z$ - $P$ relation, all superconductor under pressure will display "softening" and "hardening" behaviours depending on the sample thickness like those in Figure 4. In general, thinner samples yield higher $T_c$ at higher pressure due to "hardening". It is possible the hardening behaviour by itself could deliver further 15% increase in $T_c$, which is sufficient to raise the current experimentally confirmed $T_c$ of $LaH_{10}$ around 250-260 K [5,6] to 287.5-299 K, or the room temperature range of 14.3-25.9 °C [2,3]. The linear $T_c$ - $P$ relation described by Eq. (7) has been confirmed by all the measurements in Figure 3(b) and Figure 4 and more in [2] in the initial increasing phase. Both "hardening" (with higher $T_c$ and $P$) and "softening" behaviours can be clearly seen in Figure 4. The result of 260 K appears to be in the "hardening" stage.

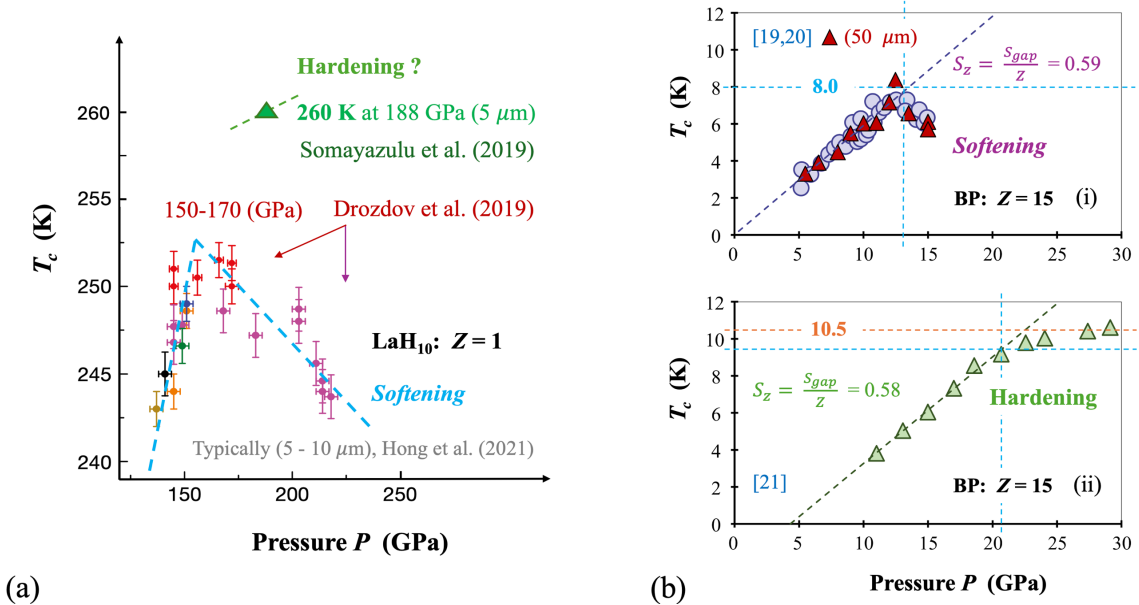

FIG. 4. (a) Two separate $LaH_{10}$ measurements [5,6]. One set (without thickness information) shows a clear "softening" behaviour after the initial linear $T_c$ - $P$ relation. The other set [6] specified the initial La metal plate was 5 micron in thickness. (b) Black phosphorus superconductor: Only one group [19] of two "softening" measurements specified the thickness of 50 microns. The "hardening" $T_c$ - $P$ relation [21] did not specify the thickness. In general, higher $T_c$ and "hardening" are linked to thinner samples. The $T_c$ increase from 8 to 10.5 K (softening to hardening) is over 30%.

Table 1 shows the thickness effect on $T_c$ of cuprate B1223 superconductor even though the thickness variation is limited. Normally, to show a clear thickness effect, the thickness variation with a factor of 10 is preferred, e.g., from 8 to 80 microns. The highest $T_c$ from the 25 $\mu$m thickness group is linked to the highest $P$, showing the "hardening" characteristics. Since 7% enhancement from 153 to 164 K has already been observed as shown in Table 1, around 15% enhancement in $T_c$ from 153 K could be expected if the sample thickness is reduced to 8 microns. Note that 15% increase from 260 K in Figure 4(a) will reach to 299 K or 26 °C.

Table 1: Thickness effect on $T_c$ of cuprate B1223 showing 7% enhancement from 153 to 164 K

| Sample thickness (µm) | $T_{c\text{-}max}$ (K) | $P$ (GPa) | Reference |
|---|---|---|---|
| 25 | 164 | 23 – 30 | [22] |
| 30 | 157 | 23 | [23] |
| 80 | 153 | 10 – 15 | [24] |

Science in 2021 [25] has listed the high-temperature superconductivity mechanism as a key challenge for future exploration and discovery. It has also been commented in [26] that "external pressure can enhance the maximum temperature of superconductivity $T_c$ in excess of what can be achieved by chemistry has puzzled researchers for decades". The macroscopic quantum tunnelling methodology summarised in Figure 1 and 2 has provided tentative answers to those two challenging questions. Importantly, we have shown the idealised positive lattice model for the free flow of Cooper pairs as illustrated in Figure 1(b) is a correct simplification since it can be generated from a lattice model with electron clouds under extreme pressure, as illustrated in Fig. 1(c).

Finally, it seems that the uniaxial pressure (not hydrostatic pressure assumed inside DAC) is preferred for creating the "low-electron-density" superconductive channels in hydrides. Thinner samples can induce the "hydrostatic" to "uniaxial" pressure transition in DAC [3], which is certainly critical to RT-S measurements. It should also be mentioned that within four months after the initial thickness effect paper was put online [2], RT-S was successfully measured by a separate group using ultra-thin hydride samples around 1 micron in thickness [27]. With further reduction in the barrier width (the distance between the tips of metal probes) for optimum quantum tunnelling, RT-S of super-hydrides [28] is virtually guaranteed. For more than a century, "room-temperature superconductivity is arguably the greatest challenge in condensed matter physics" [1]. Taking the view of quantum tunnelling, this study has shown that reduction in both the barrier height (from the formation of superconductive channels under extreme pressure) and barrier width (from utilizing the size effects for optimum electron tunnelling) is critical to the realization of RT-S.